# Feature reduction for machine learning on molecular features: The GeneScore

Alexander Denker[1,*], Anastasia Steshina[1,*], Theresa Grooß[1], Frank Ückert[1], Sylvia Nürnberg[1,§]

[1]Medical Informatics for Translational Oncology, German Cancer Research Center (DKFZ), Heidelberg, Germany

* equal contribution

§ corresponding author
Email: s.nuernberg@dkfz-heidelberg.de

Abstract

We present the GeneScore, a concept of feature reduction for Machine Learning analysis of biomedical data. Using expert knowledge, the GeneScore integrates different molecular data types into a single score.
We show that the GeneScore is superior to a binary matrix in the classification of cancer entities from SNV, Indel, CNV, gene fusion and gene expression data.
The GeneScore is a straightforward way to facilitate state-of-the-art analysis, while making use of the available scientific knowledge on the nature of molecular data features used.

Introduction

With recent discoveries in high-throughput biomedical technologies such as next-generation sequencing large molecular data sets have become available for downstream analyses. The availability of big data suggests the use of machine learning methods, where rather than making ad hoc assumptions based on expert knowledge the algorithm learns important features from the data itself.

However, in clinical data sets the number of features often greatly outweighs the number of observations (patients). This imbalance bears the risk of overfitting the algorithm to the data. Here, the algorithm closely describes the patient cohort of the training set but lacks transferability to new cohorts and thus fails to predict future observations. One measure to combat this lies in the reduction of features prior to analysis, such as the removal of those features with low variance.

Here, we suggest the integration of different molecular features of the same gene into one score, the GeneScore (Figure 1). The study uses single-nucleotide variants (SNV), insertions/deletions (Indels), gene fusions, copy-number variants (CNV) and gene expression of cancer patients from the MASTER study[1], a precision oncology program dedicated to the identification of novel treatment approaches based on molecular profiling.

Using expert biomedical knowledge we assign each molecular feature a score based on their predicted phenotypic effect. We then use the combined gene matrix for the classification of cancer entities using a gradient boosted tree classifier, and compare the performance to that using a binary matrix.

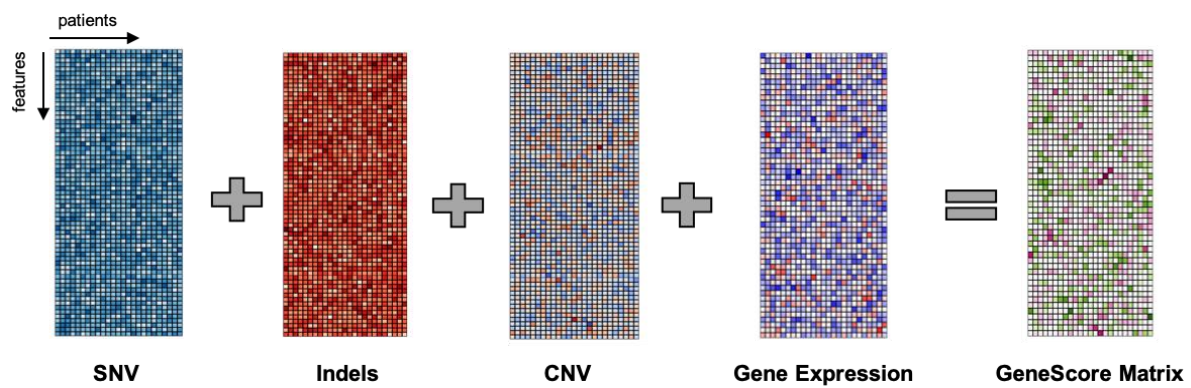

Figure 1. Composition of the GeneScore, which combines SNVs, Indels, CNVs, Fusions and gene expression in one matrix

## Materials and Methods

**Data source**

Anonymized molecular data from MASTER study participants was used after bioinformatics pre-processing using the DKFZ One-Touch Pipeline[2].

**Calculation of the GeneScore**

After identifying all patients with a full data set, each molecular data type was scored as follows (Table 1):

Table 1. Data types used for the GeneScore

| Data Type | Pre-processing Method | Scale |
| --- | --- | --- |
| SNV | CADD score | 0 – 1 |
| Indels | Exonic Classification & Confidence | 0. 5 – 1 |
| CNV | Total copy number | 0 – 1 |
| Fusion Genes | Binary | 0 / 1 |
| Expression | log2 FC against GTEx v8 mean | 0 – 1 |

SNVs were annotated using the Combined Annotation Dependent Depletion (CADD) score[3], which weighs the deleteriousness of the nucleotide change. The score was rescaled to a scale between 0 and 1.

Indels were scored using their exonic classification and confidence. A new numeric value on the scale of 0.5 to 1 was then derived from merging these two levels of information. Frameshift modifications by deletion or insertion or gain/loss of stop codon were classified as High Damage, while insertions and deletions that do not cause a frameshift were classified as Moderate Damage (Figure 2).

| Score | Confidence | Damage |
| --- | --- | --- |
| **1.0** | 10 | High |
| **0.9** | 9 | High |
| **0.8** | 8 | High |
| **0.7** | 10 | Moderate |
| **0.6** | 9 | Moderate |
| **0.5** | 8 | Moderate |

- High damage: frameshift deletion / frameshift_insertion / stopgain / stoploss
- Moderate damage: nonframeshift deletion / non-frameshift insertion

Figure 2. Scaling of the Indel score

The CNV score used total copy number (TCN) and was rescaled from 0 to 1. Fusion genes were scored on a binary scale (0 or 1).

Due to the absence of sequenced paired normal tissue, we chose to calculate differential gene expression using v8 data from the Genotype-Tissue Expression (GTEx) project. We calculated log2 fold expression against the mean of all GTEx tissue types using Limma-voom[4]. The resulting values were then rescaled to be between 0 and 1.

The GeneScore matrix was created by adding all scores per patient per gene. Therefore, the resulting GeneScore is a sum of several elements:

$$score(i,j) = cnv(i,j) + gene\ epxression(i,j) + indels(i,j) + snv(i,j) + fusions(i,j)$$
$$i=1, \ldots, N\ genes \quad j = 1, \ldots, M\ patients$$

The shape of the GeneScore matrix for the 5-way classification task was (22914, 445), where 22914 represents the number of features/genes and 445 the number of observations/patients.

All non-informative genes (all values 0) were removed from the GeneScore matrix.

**Binary Score**

In order to demonstrate that the idea of the GeneScore has merit, we compared the performance of the classification model using GeneScore to the binary score. The binary matrix was created by modifying the GeneScore matrix so that all values >0 were set to 1. This way, a resulting matrix had the same number of genes as the GeneScore matrix, but a single alteration on any level (SNV, CNV, Indels or Fusions) would equally lead to the maximum change of the feature's value.

**Model Training**

Scikit-learn[5], a popular Python module for predictive data analysis, was used for machine-learning algorithm creation. Cancer patient classification tasks were performed using eXtreme Gradient Boosting (XGBoost), an optimized distributed gradient boosting decision tree algorithm. An ensemble approach of weak decision trees was chosen for enhanced final prediction. The selected model used a gradient descent algorithm to minimize the loss when adding new models. It was run with two options – with or without Recursive Feature Elimination (RFE). For evaluation of feature selection stability the analysis were run multiple times.

Table 2: List of Python modules used for training the GeneScore models

| Module | Version |
|---|---|
| pandas | 1.1.0 |
| numpy | 1.19.1 |
| sklearn | 0.23.1 |
| xgboost | 1.1.1 |
| matplotlib | 3.3.0 |

For upsampling Synthetic Minority Over-sampling (SMOTE)[6] was applied to the training data, while no upsampling was performed on test data. Train-test split followed a 70/30 distribution.

Results

**Characteristics of the GeneScore**

The range of the resulting GeneScore matrix values was between 0 and 3.84. The matrix had a sparsity of around 12% (Figure 3).

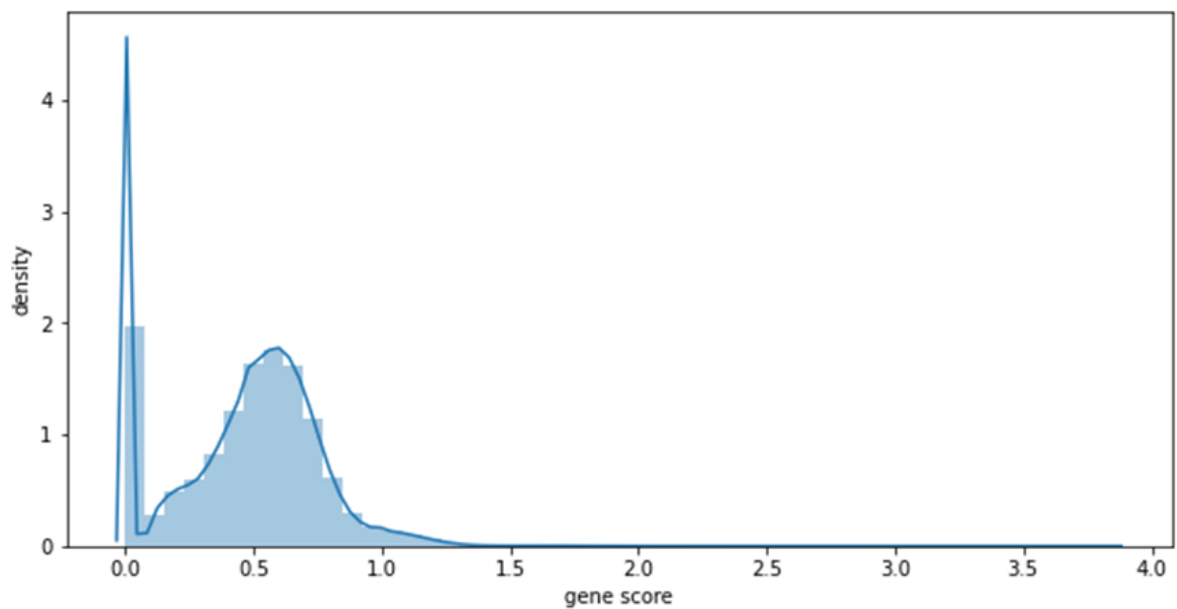

Figure 3. Distribution of the GeneScore

Examining the contributions of the different data types to the GeneScore shows that expression is the data type that contributes the most to the score, whereas the other data types only contribute relatively little (Figure 4).

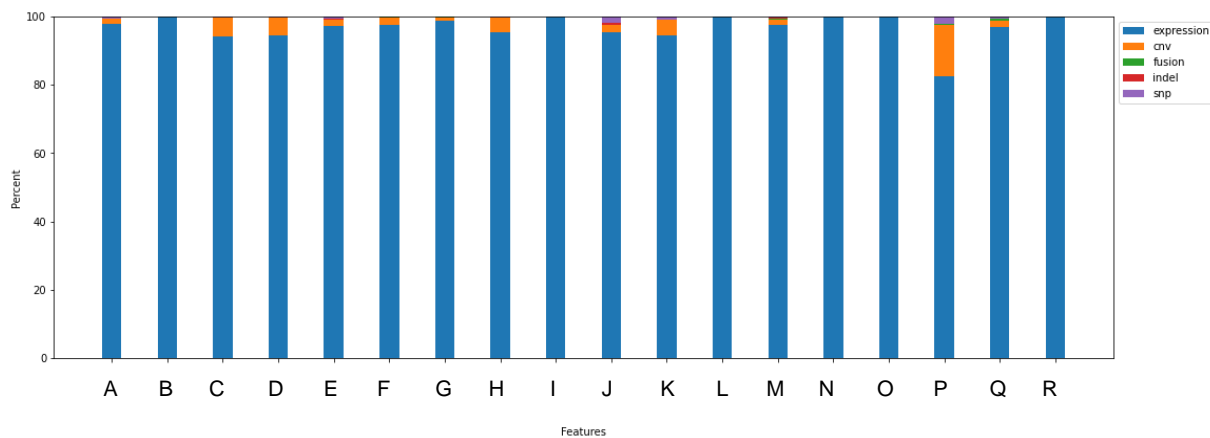

Figure 4. GeneScore contributions of different data types for top features from the classification on five cancer entities

The GeneScore values display different distributions across the different cancer entity classes as seen here for top features of a five-class classification task

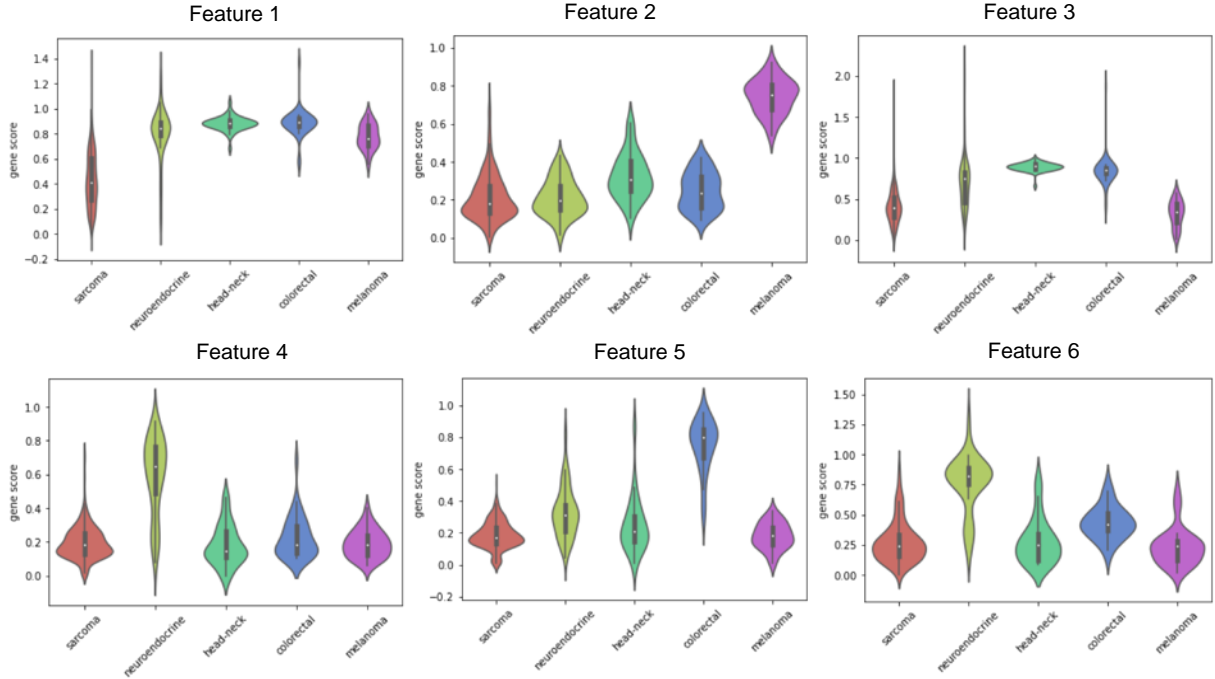

Figure 5. Violin plots of features as examples of GeneScore distributions across five different cancer entities

Evaluated for the 100 features with the highest standard deviation across all patients, the GeneScore shows distinct clusters of high values, which do however not correspond to the cancer entity label (Figure 6).

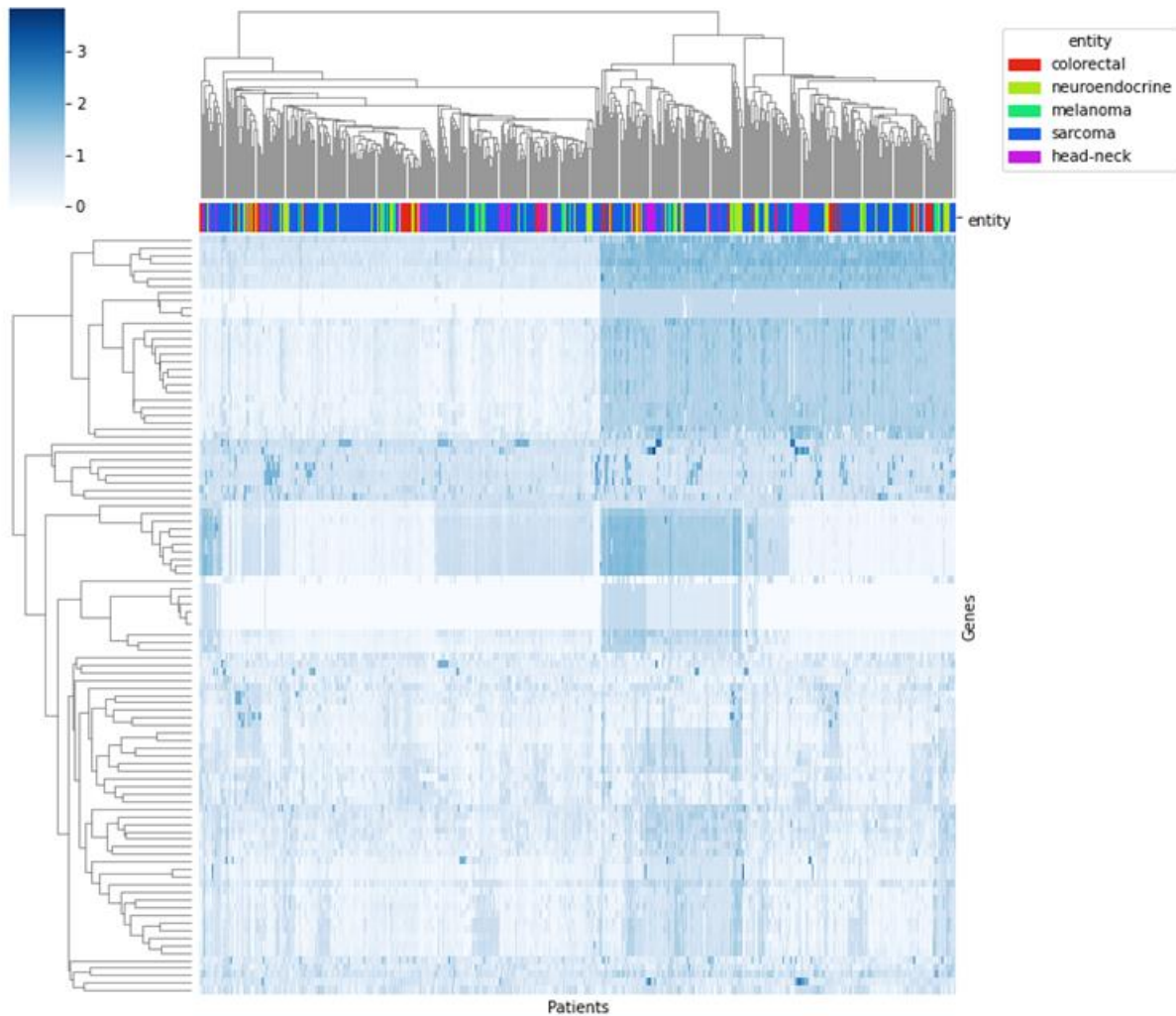

Figure 6: Clustering of GeneScore values for 100 most variable genes across all cohort patients

**Performance of the GeneScore in a binary classification task**

The MASTER cancer cohort comprised of 779 patients of which 278 were Sarcoma patients. For this analysis the Sarcoma patients were assigned label '1', while other entities were assigned label '0'.

We compared the performance of the GeneScore to a binary matrix derived from the GeneScore matrix by reassigning all non-zero values to 1. We performed the analysis either with or without Recursive Feature Elimination (RFE) and compared standard performance metrics (Table 3).

Table 3: Performance criteria of 4 different binary classification approaches

|  | A | B | C | D |
|---|---|---|---|---|
| Matrix | GeneScore | Binary | GeneScore | Binary |
| RFE | No | No | Yes | Yes |
| Train: Accuracy | 1.000 | 0.996 | 1.000 | 0.928 |
| Test: Accuracy | 0.906 | 0.697 | 0.885 | 0.598 |
| Test: AUC | 0.965 | 0.710 | 0.956 | 0.602 |
| Test: Precision (weighted average) | 0.878 | 0.614 | 0.860 | 0.421 |
| Test: Recall (weighted average) | 0.857 | 0.416 | 0.809 | 0.321 |
| F1 (weighted average) | 0.867 | 0.496 | 0.834 | 0.364 |

This illustrated that both analyses using the GeneScore matrix show a substantially better performance in all test criteria (see also Figure 7).

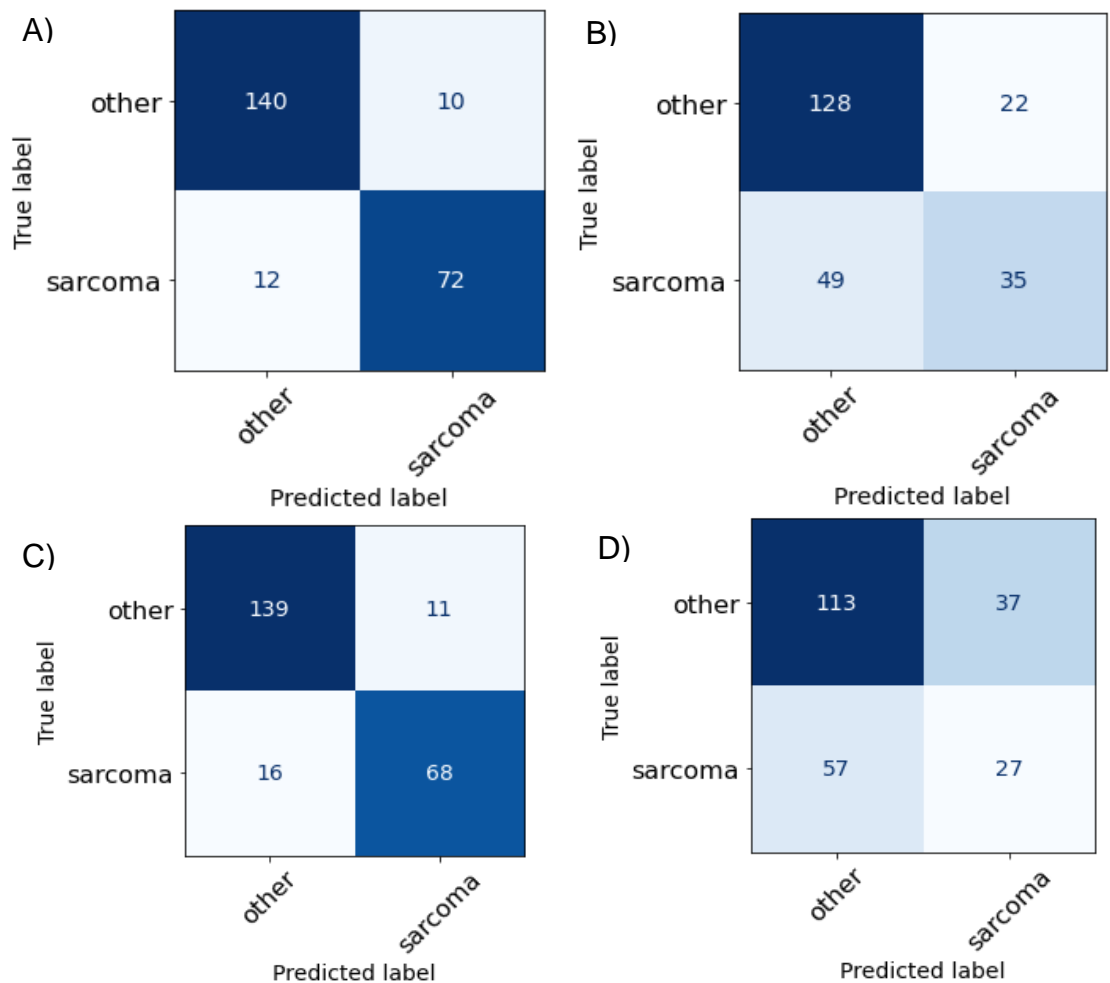

Figure 7: Confusion matrices of binary classifications for A) GeneScore without RFE, B) Binary without RFE, C) GeneScore with RFE, D) Binary with RFE

Since the two cancer entitles groups, sarcoma and others, were highly imbalanced, we additionally evaluated if upsampling had an effect on the performance of the GeneScore classification model (without RFE). However, this was not the case: Upsampling with SMOTE did not lead to an increase in accuracy, AUC, precision or F1 score (Table 4).

Table 4. Results demonstrating that oversampling did not improve model accuracy

|  | Without Upsampling | With Upsampling |
|---|---|---|
| Train: Accuracy | 1.00 | 1.00 |
| Test: Accuracy | 0.933 | 0.92 |
| Test: AUC | 0.974 | 0.972 |
| Test: Precision | 0.90 | 0.86 |
| Test: Recall | 0.913 | 0.925 |
| F1 | 0.91 | 0.89 |

**Performance of the GeneScore in a five-way classification task**

The results of the cancer entity classification without RFE show that the label predicted by the classification and the true label largely overlap, with some discrepancies. The majority of the patients were Sarcoma patients, of which 79 were identified correctly and 1 was misattributed to the neuroendocrine class. Of the other 54 samples, 41 were classified correctly and 13 were misidentified (Figure 8A). Overall, the mean accuracy was 0.903 with a mean weighted F1 score of 0.89.

Using RFE in the classification task leads to similar results to the classification without RFE with a large overlap between true and predicated samples, but also some discrepancies. Again, 79 Sarcoma patients were identified correctly and 1 was wrongly attributed to the neuroendocrine class. Of the other 54 samples, 42 were classified correctly and 12 incorrectly (Figure 8B). Overall, the mean accuracy was 0.903 with a mean weighted F1 score of 0.90.

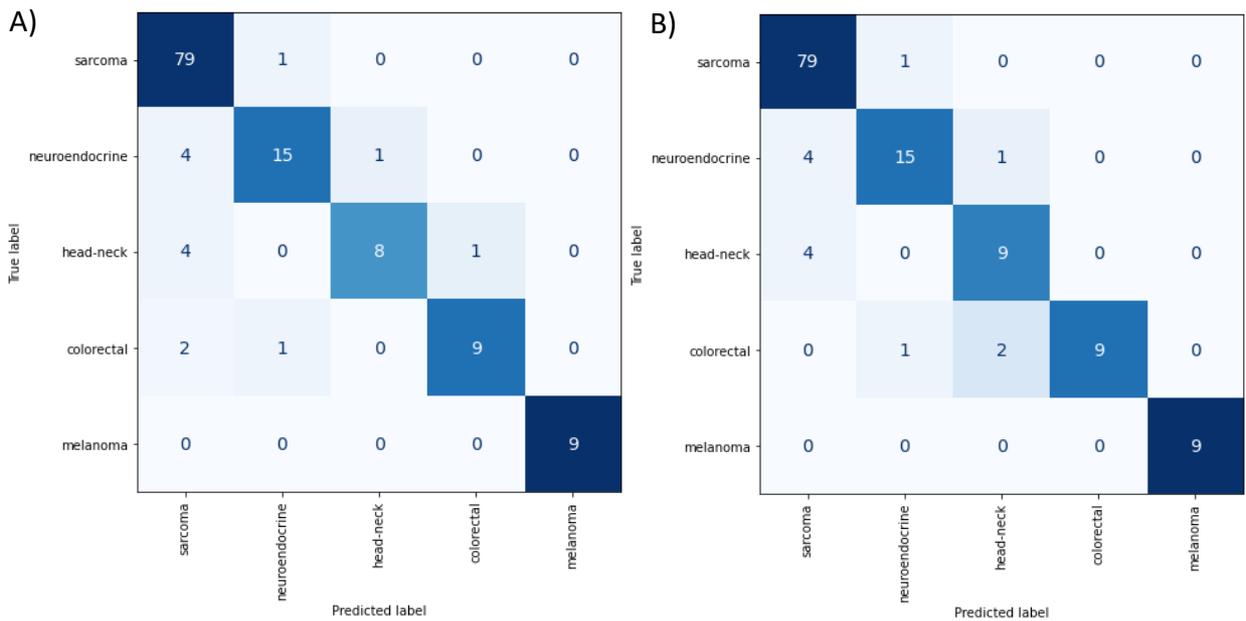

Figure 8. A) Results of cancer entity classification using XGBoost run 100 times without RFE or B) with RFE.

To evaluate the stability of feature selection we ran each analysis multiple times and compared the times selected, mean feature importance and standard deviation of the top 25 features based on their mean feature importance. Figure 9 illustrates that there is great variability in the feature importance of each feature across runs, leading to a large standard deviation over the mean. The analysis including RFE also shows that important features are not selected during each analysis run.

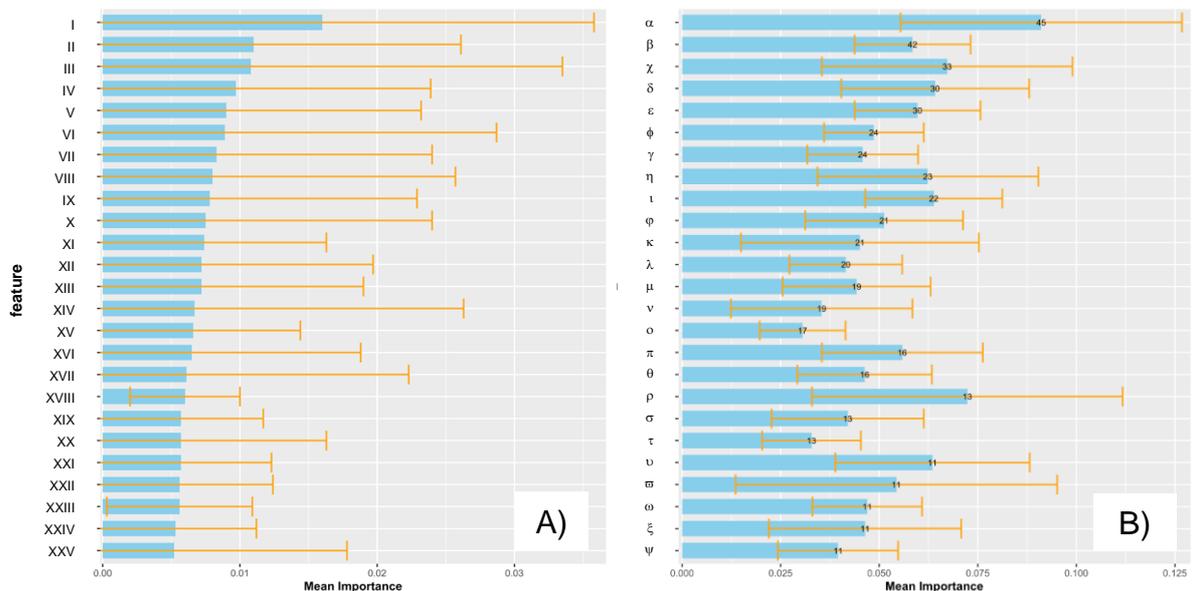

Figure 9: Mean feature importance, standard deviation and times selected (for analysis with RFE only) from analysis run 50 times A) without RFE or B) with RFE

Conclusions

Using the GeneScore for the integration of molecular data is superior to a binary matrix in a gradient boosted tree classification task. In all tests, the GeneScore performed better, with higher accuracy, AUC, precision and recall than the binary score, demonstrating that the principle of the GeneScore has merit. The resulting important features resulting from the analysis can be readily used for feature annotation, and pathway or gene set enrichment analysis using public databases and bioinformatics tools in R/Bioconductor.

References

1. Horak, P. *et al.* Comprehensive genomic and transcriptomic profiling in advanced-stage cancers and rare malignancies: Clinical results from the MASTER trial of the German Cancer Consortium. *Annals of Oncology* **30,** vii24; 10.1093/annonc/mdz413.085 (2019).
2. Reisinger, E. *et al.* OTP: An automatized system for managing and processing NGS data. *Journal of biotechnology* **261,** 53–62; 10.1016/j.jbiotec.2017.08.006 (2017).
3. Rentzsch, P., Witten, D., Cooper, G. M., Shendure, J. & Kircher, M. CADD: predicting the deleteriousness of variants throughout the human genome. *Nucleic Acids Research* **47**; 10.1093/nar/gky1016 (2019).
4. Law, C. W., Chen, Y., Shi, W. & Smyth, G. K. voom: Precision weights unlock linear model analysis tools for RNA-seq read counts. *Genome Biology* **15**; 10.1186/gb-2014-15-2-r29 (2014).
5. Fabian Pedregosa *et al.* Scikit-learn: Machine Learning in Python. *Journal of Machine Learning Research* **12,** 2825–2830 (2011).
6. N. V. Chawla, K. W. Bowyer, L. O. Hall & W. P. Kegelmeyer. SMOTE: Synthetic Minority Over-sampling Technique. *1* **16,** 321–357; 10.1613/jair.953 (2002).